\documentclass[sigconf]{acmart}

\usepackage{graphicx}
\usepackage{colortbl}
\usepackage{balance}
\usepackage{subcaption}
\usepackage{booktabs}
\usepackage{multirow}

\AtBeginDocument{%
  }

\copyrightyear{2025}
\acmYear{2025}
\setcopyright{rightsretained}
\acmConference[CHIWORK '25]{CHIWORK '25: Proceedings of the 4th Annual Symposium on Human-Computer Interaction for Work}{June 23--25, 2025}{Amsterdam, Netherlands}
\acmBooktitle{CHIWORK '25: Proceedings of the 4th Annual Symposium on Human-Computer Interaction for Work (CHIWORK '25), June 23--25, 2025, Amsterdam, Netherlands}
\acmDOI{10.1145/3729176.3729183}
\acmISBN{979-8-4007-1384-2/25/06}

\begin{document}

\title[Effects of Avatar Representation in Video-Mediated Collaborative Interactions]{Perception in Pixels: Effects of Avatar Representation in Video-Mediated Collaborative Interactions}

\author{Pitch Sinlapanuntakul}
\email{wspitch@uw.edu}
\orcid{0000-0003-3551-8531}
\affiliation{%
  \institution{Human Centered Design and Engineering \\ University of Washington}
  \city{Seattle}
  \state{Washington}
  \country{USA}
}

\author{Mark Zachry}
\email{zachry@uw.edu}
\orcid{0000-0002-1067-7168}
\affiliation{%
  \institution{Human Centered Design and Engineering \\ University of Washington}
  \city{Seattle}
  \state{Washington}
  \country{USA}
}

\renewcommand{\shortauthors}{Sinlapanuntakul and Zachry}

\begin{abstract}
  Interactive collaborative video is now a common part of remote work. Despite its prevalence, traditional video conferencing can be challenging, sometimes causing social discomforts that undermine process and outcomes. Avatars on 2D displays offer a promising alternative for enhancing self-representation, bridging the gap between virtual reality (VR) and traditional non-immersive video. However, the use of such avatars in activity-oriented group settings remains underexplored. To address this gap, we conducted a mixed-methods, within-subject study investigating the impacts of avatar-mediated versus traditional video representations on collaboration satisfaction and self-esteem. 32 participants (8 groups of 4 with pre-established relationships) engaged in goal-directed activities, followed by group interviews. Results indicate that avatars significantly enhance self-esteem and collaboration satisfaction, while qualitative insights reveal the dynamic perceptions and experiences of avatars, including benefits, challenges, and factors influencing adoption likelihood. Our study contributes to understanding and implications of avatars as a camera-driven representation in video-mediated collaborative interactions.
\end{abstract}




\maketitle

\section{Introduction}

A broad social shift toward remote work using synchronous video-based platforms has drawn attention to the fundamentals of collaborative interactions in group settings. However, working together on these platforms often introduces challenges associated with self-representation and well-being. For many, being on camera induces social discomfort and self-awareness, undermining self-esteem and diminishing active participation. Such effects, akin to ``Zoom anxiety'' \cite{falzone2005, ramasubramanian2020}, hinder collaboration satisfaction, which is a critical factor in effective teamwork. To mitigate this, individuals may turn off their cameras, often at the cost of reduced visual presence and perceived engagement, which in turn weakens group dynamics and collaborative outcomes \cite{asher2020, azriel2020, lin2021, vriends2017}.

One promising approach to improving self-esteem and collaboration satisfaction involves the use of avatars. Research in immersive VR environments has demonstrated significant benefits of avatars, such as enhanced social presence through embodied nonverbal communication, reduced anxiety through customizable self-presentation, and improved collaborative task performance \cite{freeman2021, freeman2022, tarnec2023}. Despite these promising findings in VR contexts, the effects of using avatars in non-immersive video meetings—the primary medium across social, educational, and professional contexts—remains largely unexplored. Our work addresses this gap by examining how avatars can transform standard video conferencing experiences. By replacing portions of live video feeds with expressive, customizable representations, avatars may create a more engaging and less self-conscious environment while maintaining essential interpersonal connections. This approach offers a potential solution to Zoom anxiety without the heightened self-awareness inherent in traditional video meetings.

As technology continues to shape the way we engage and interact in digitally-mediated workspaces, the emerging and evolving role of avatars, as camera-driven representations, has become increasingly relevant. While an audio-only modality can reduce Zoom anxiety, it can lead to feelings of exclusion from the group \cite{panda2022}. Avatars, offering a middle ground between audio and video representation, augment the physical self to enhance presence while addressing the challenges associated with live video interaction \cite{park2023, ratan2022, wicks2021}. Unlike VR avatars that obscure subtle cues like head movements and body language, avatars in 2D video conferencing retain essential aspects of live-streamed communication while providing an alternative to full video exposure \cite{nowak2018}. Park et al. \cite{park2023} suggests the positive impact of Memoji avatars on participation in one-on-one debates when paired with self-affirmation techniques, with the potential downsides, such as impaired response to critical feedback. This then raises questions about the role of avatars in group-based, activity-oriented collaborative settings.

Consequently, in this paper, we explore the use of avatars for activity-driven collaborations in 2D video interfaces such as mobile or laptop screens, moving beyond prior studies focused on VR environments or non-task-oriented discussions. We conducted a mixed-methods study, combining a within-subject experimental design with group interviews (\textit{N} = 32; 8 groups of 4 participants each) to answer the following research questions (RQ):

\begin{itemize}
    \item \textbf{RQ1}: How does the use of avatars in video-mediated collaborative interactions impact self-esteem and collaboration satisfaction compared to traditional video representations?
    \item \textbf{RQ2}: In what ways do avatars influence the perceptions of self and others in video-mediated collaborative interactions?
    \item \textbf{RQ3}: How do shared experiences with avatars in video meetings shape the perceived appropriateness and potential future use of this technology in collaborative settings?
\end{itemize}

Our findings indicate that, compared to traditional video meetings, avatars significantly lead to increased levels of self-esteem and collaboration satisfaction in video-mediated collaborative interactions. When using avatars, participants reported higher self-esteem, attributed to the avatars' ability to reduce self-discrepancies, enable customization, and increase a sense of control. Moreover, collaboration satisfaction improved across multiple dimensions, including efficiency, coordination, fairness, understanding, and trust in group decisions. From group interviews, participants noted both benefits (e.g., increased self-confidence, engaged participation, comfort of physical self, and reduced social tension) and challenges (e.g., emotional expression and interpretation), reflecting current technological limitations, when using avatars. We find that avatars influence not only individual self-perception and experience but also group dynamics in collaborative video-mediated interactions. The core contributions of our work are: (1) early empirical evidence on the impact of avatars in group-based, activity-oriented video-mediated interactions and (2) implications and insights into the benefits, challenges, and factors influencing avatar use in non-immersive 2D video meetings.

\section{Background and Related Work}

Our study of the impacts of avatar representation in video-mediated collaborative interactions has connections to two bodies of research: avatar-mediated communication in videoconferencing and the effects of avatars on self-esteem.

\subsection{Avatar-Mediated Communication in Videoconferencing}

While avatar-mediated communication has been extensively researched in immersive VR environments, its application in remote collaboration, especially within non-immersive interfaces, remains underexplored. Most HCI research on avatars in VR is confined to controlled laboratory settings, potentially limiting their real-world applicability \cite{aufegger2022, dobre2022, hudson2016, phadnis2023, yoon2019}. The limited accessibility of VR devices further restricts their application to specialized domains, largely in training and skill development \cite{bartl2022, grassini2020, xie2021}, which may not fully reflect the experiences of broader populations. Over the past decade, significant attention has been given to group interactions using avatars in VR environments, covering factors such as transferability, performance, presence, co-presence, and trust \cite{moustafa2018, pan2017, schroeder2002, wu2021}. These findings, however, cannot be directly translated to the 2D, screen-based nature of video meeting platforms. For instance, Bente et al. \cite{bente2008} compared text-chat, audio, and video conferencing with avatars, demonstrating distinctions between text-based and real-time modalities but not among audio, video, and avatars-mediated formats. In work settings, Inkpen and Sedlins \cite{inkpen2011} observed that people were most receptive to avatars that were most like webcam photos.

In contrast to VR avatars, screen-based avatars, facilitated by smartphone or web camera-based applications, remain crucial due to their widespread availability and ease of use \cite{bonfert2023, makled2022}. Most existing research focuses on immediate and context-specific objectives, such as participation in online social events and team-building activities. Screen-based avatars offer several advantages. For instance, Liu et al. \cite{liu2020} revealed that avatars delivered a more realistic representation of garment try-outs as compared to their VR counterparts. Another study demonstrated reduced mental workloads when users interacted with information through screen-based avatars compared to those using VR headsets \cite{li2021}. This suggests that avatars can offer a more comfortable and less cognitively demanding experience. Furthermore, Zibrek et al. \cite{zibrek2018} found that avatars, when designed with sufficient visual and behavioral fidelity, were just as appealing and engaging as their VR counterparts, indicating their effectiveness in eliciting positive user interactions.

Although no research prior to ours has specifically explored group collaboration with avatars in videoconferencing, a few prior studies have explored the use of avatars for realism, presence, and self-affirmation for camera shyness. A recent large-scale survey study found that avatar realism significantly influenced acceptability for videoconferencing among knowledge workers, with less realistic avatars being more acceptable for known colleagues but less so for managers and unknown colleagues \cite{phadnis2023}. Critical attributes included professionalism, credibility, and seriousness, though some participants valued the added benefits of cuteness, fun, and liveliness when using avatars at work. Panda et al. \cite{panda2022} investigated the impact of low-fidelity, non-customizable avatars on presence and co-presence in mixed-modality conferencing, including audio-only, traditional video, and avatar representations. They found that participants preferred avatars representing themselves and others when there is a lack of visual representation (i.e., audio-only), which enhanced perceived presence and co-presence. Another recent study showed that using avatars in videoconferencing, when used in conjunction with self-affirmation, increased active participation in one-on-one debates but also negatively impacted how people responded to critical messages from others \cite{park2023}. 

While not directly related, a recent research on AR face filters in public speaking \cite{leong2023} provides valuable insights that can be extended to avatar-mediated communication. They suggest that AR filters as a mediator can alleviate social anxiety and enhance comfort in collaborative settings, potentially improving group participation, engagement, and experience. Even minimal animations from avatars, like subtle facial movements, contribute to users’ identification and embodiment with them, enhancing the overall experience, engagement, and behavior \cite{gonzalez2020facial}. To the best of our knowledge, no prior study has taken the initiative to design and/or empirically investigate the complexities associated with the use of avatars for collective activity achievement within video-based group settings.

\subsection{Avatars and Self-Esteem}

Research on self-esteem, the way individuals value and perceive themselves \cite{blascovich1991}, within avatar-mediated communication is limited \cite{koek2024, watts2016}. However, related concepts like self-perception have been extensively studied. Understanding how these avatars influence self-perceptions and, by extension, self-esteem has become increasingly prevalent in various forms of online interaction. Early computer-mediated communication research suggested that the relative anonymity and reduced social cues in digital environments could lead to deindividuation and altered self-perception \cite{walther2011, walther2015}. However, with the advent of more sophisticated avatars, nuanced effects on self-esteem emerge. The Proteus effect \cite{yee2007} posits that individuals may conform to the behavior and attitudes they believe others would expect from the avatar they embody. This effect has significant implications for self-esteem, as users may internalize the perceived characteristics of their digital representations. Embodying, for example, taller avatars can lead to more confident behavior in negotiation, while attractive avatars can enhance self-esteem and social behavior \cite{fokides2021, fox2013, kocur2020, messinger2008, nowak2018, praetorius2020}. 

The malleability of self-representation in digital environments allows users to experiment with different identities, potentially leading to both positive and negative impacts on self-esteem. Research has demonstrated that avatar customization can serve as a form of self-expression and identity exploration, often resulting in improved self-esteem when users create idealized versions of themselves \cite{dunn2012, ratan2014, sibilla2018}. Conversely, the discrepancy between one’s avatar and real-life appearance can sometimes lead to decreased body satisfaction and self-esteem, particularly when the idealized digital representation is perceived as unattainable \cite{cacioli2014, park2018}. The relationship between avatar embodiment and self-esteem is further complicated by the social context of the interaction. In collaborative settings, the perception of one’s avatar by others can significantly influence self-esteem. Positive feedback on one’s avatar can lead to increased self-esteem, while negative social experiences can have detrimental effects \cite{lemenager2020, nowak2018, praetorius2020}.

Recent research has extended findings on embodiment and self-esteem in avatar-mediated communication to augmented reality (AR) filters, showing similar patterns of internalization and self-perception shifts among young adults \cite{sinlapanuntakulandzachry2024}. For example, frequent use of a puppy AR filter was associated with a desire to be perceived as cute and playful, mirroring the filter's characteristics \cite{fribourg2021, sinlapanuntakulandzachry2024}. This aligns with observations in avatar-mediated environments, where self-esteem is influenced by the attributes of digital representations \cite{fokides2021, messinger2008, praetorius2020}. Avatars, here, present a unique case where the digital elements are displayed on 2D, non-immersive displays like mobile phones, unlike VR that provides fully immersive experiences or AR that superimposes 3D information onto the user’s physical environment \cite{sinlapanuntakul2023, sinlapanuntakul2022}. This form of mediated interaction raises questions about their influence on self-perception and self-esteem. Research on immersive VR has shown that highly personalized avatars, mapped to users’ facial features, significantly impact emotions, body satisfaction, and feelings of presence \cite{dubosc2021, freeman2021, hartbrich2023, ohkruzic2020, waltemate2018}. However, the specific effects on self-esteem in video-based collaborative remain largely unknown. Avatars hold a unique opportunity to enhance nonverbal communication while maintaining a pure form of interaction. Understanding their impact on self-esteem is crucial, as it profoundly influences collaboration dynamics, social behavior, and self-perception in mediated communication \cite{harris2020}.

\section{Methods}

\subsection{Study Design}

We used a mixed-methods approach to investigate the effects of avatar use in video-mediated collaborative interactions with goal-directed group activities on self-esteem and collaborative satisfaction. Employing a within-subject experimental design, we compared two camera-driven representation conditions: \textit{avatar} and \textit{traditional video}. Our dependent variables included self-esteem, collaboration satisfaction, and qualitative insights from post-experiment group interviews. In this context, avatars served as 2D animated representations (e.g., Memoji\footnote{https://support.apple.com/en-us/111115}) overlaid on participants’ heads within a non-immersive video interface (i.e., FaceTime), whereas traditional video pertained to direct real-time streaming of participants’ physical appearances during video meetings. The order in which we presented the conditions was counterbalanced to mitigate potential order effects across participant groups.

\begin{figure}[H]
  \centering
  \includegraphics[width=\linewidth]{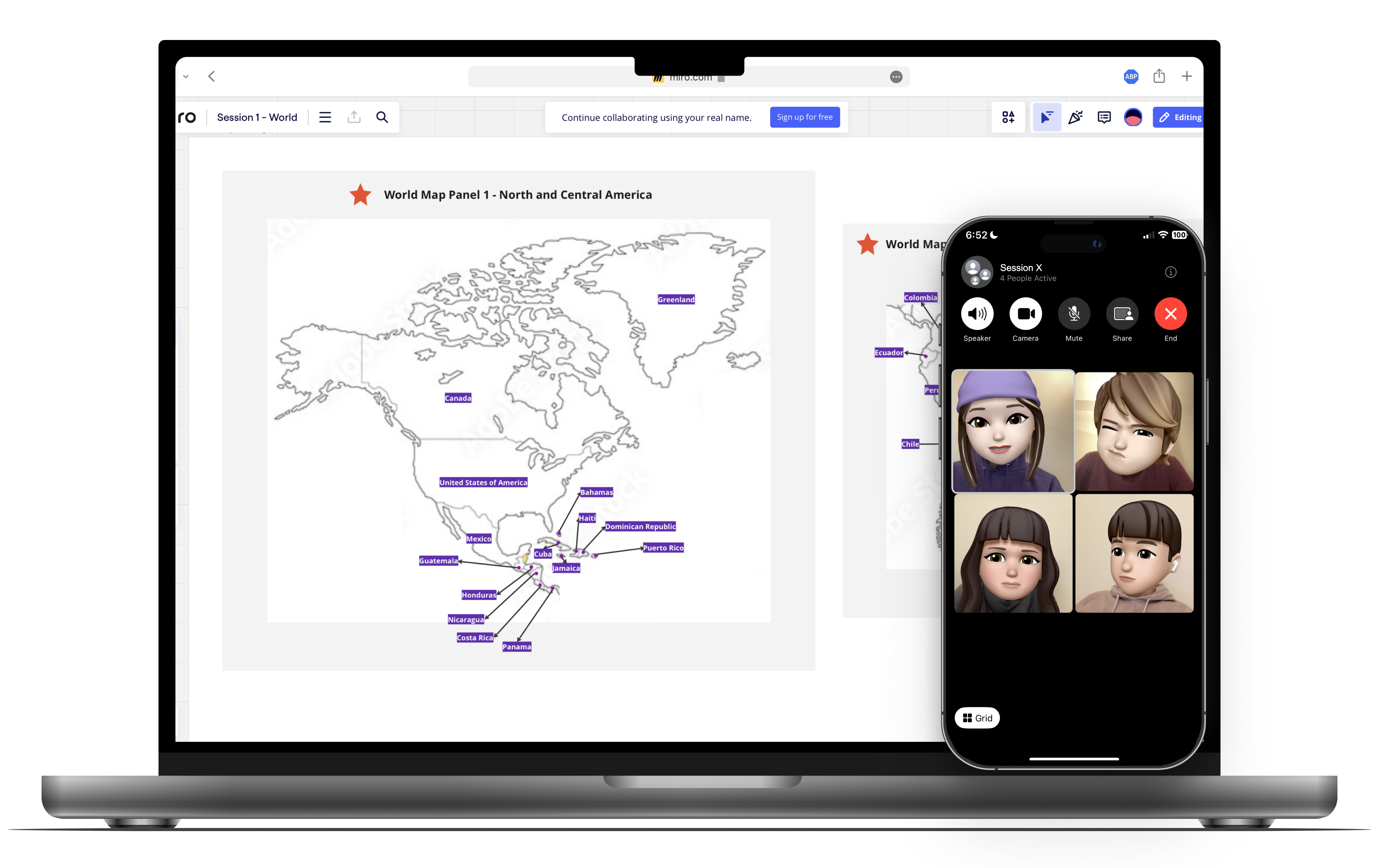}
  \caption{Device Setup with Miro (on a laptop) and FaceTime (on a mobile device).}
  \Description{The experimental setup of Miro, featuring the world map, on a laptop screen and FaceTime, featuring Memoji avatars as the camera-driven representation.}
  \label{fig:setup}
\end{figure}

Our study specifically leveraged a hybrid multi-device setup that integrates avatars (i.e., Memoji) via FaceTime on mobile devices alongside Zoom on laptops using screen sharing (Figure \ref{fig:setup}). This configuration reflects an increasingly common practice in various remote collaboration wherein individuals multitask by relying on a computer for primary tasks and secondary screen such as their mobile device for auxiliary activities like messaging, note-taking, or accessing additional resources \cite{ansah2022,son2023,suh2018}. Our study design, using a mobile device for live avatar representation, enabled participants to maintain a distinct focus on their avatar as a dynamic, personal representation while also maintaining their physical environment visible in the background, something that is less easily achieved in a computer-only setup.

\subsection{Avatar: Memoji}

In our study, avatars appeared on 2D displays, such as mobile or laptop screens, similar to traditional video meetings. Instead of live video, participants’ head movements and facial expressions were tracked and represented through their avatars. These avatars are not rendered into the real world but are presented in a standard grid layout, which maintains a familiar format for video calls and are comparable to traditional video meetings.

\begin{figure}[H]
    \centering
    \includegraphics[width=0.94\linewidth]{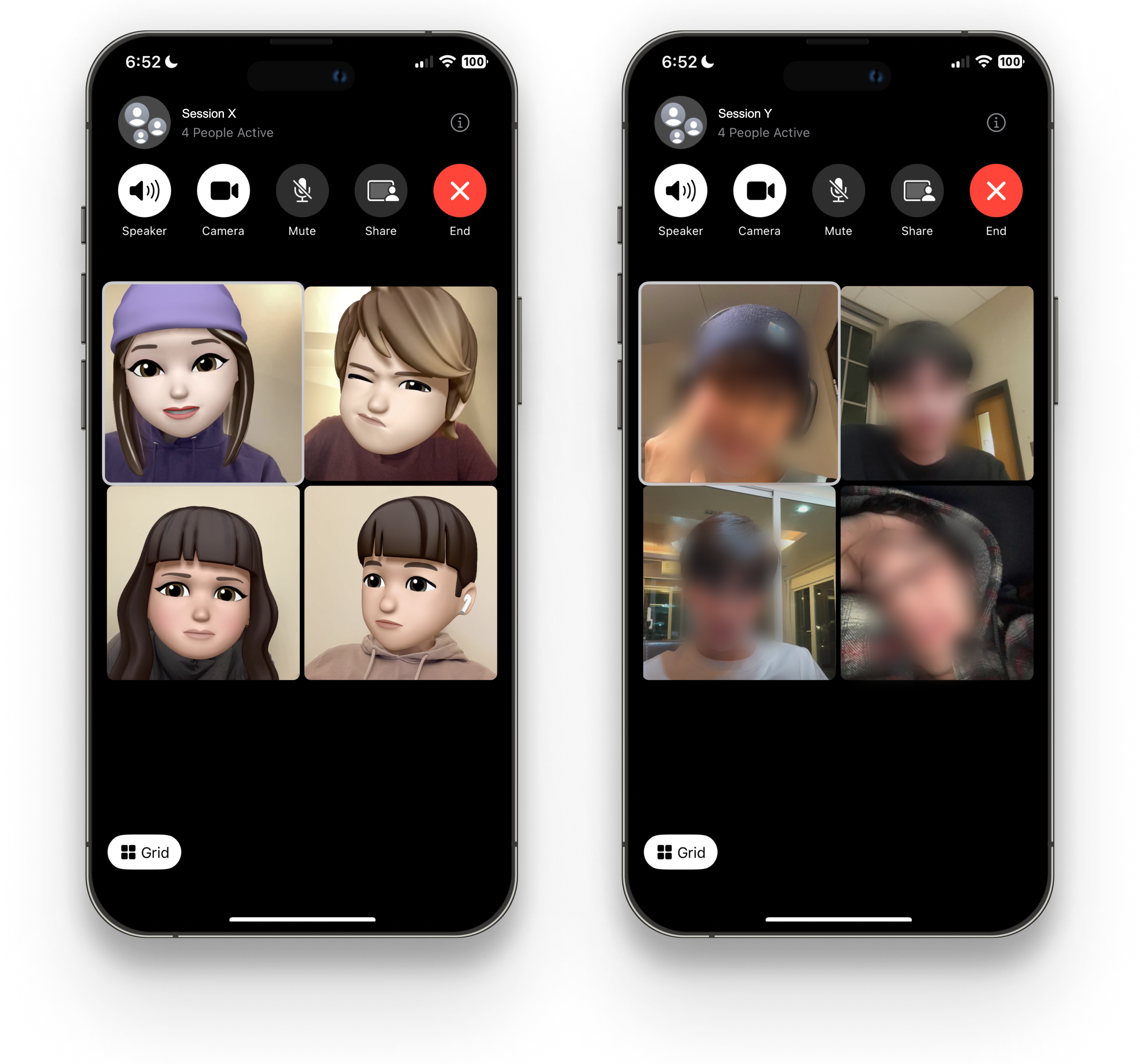}
    \caption{Avatar condition (left) vs. video condition (right) in a 2x2 grid on FaceTime. Example participant representations are blurred to maintain anonymity and privacy.}
    \Description{(Left) 2x2 grid FaceTime on mobile with each grid featuring a Memoji avatar. (Right) 2x2 grid FaceTime on mobile with each grid featuring a participant. Participants' faces are blurred for anonymity and privacy.}
    \label{fig:avatar-vs-video}
\end{figure}

During our data collection period in mid-to-late 2023, Apple’s Memoji technology emerged as the most advanced and accessible screen-based humanoid avatar option, making it particularly well-suited for the friend-based interactions central to our study. While platforms such as Zoom, Microsoft Teams, and Google Meet had begun incorporating avatar features, these either lacked the humanoid characteristics or replaced both the user and their background with entirely 2D visual elements. In contrast, Memoji on FaceTime uniquely altered only the user's head while retaining the surrounding physical environment on-screen (Figure \ref{fig:avatar-vs-video}), thus providing a more authentic and familiar experience for video communication \cite{suda2021}.

\subsection{Design of Game-Based Maps and Activities}
\label{map-design}

We thoroughly designed two maps (Figure \ref{fig:maps}), a world and a university campus maps, on Miro\footnote{https://miro.com/} as a central element in structured game-based activities. Following our UX design approach in Sinlapanuntakul et al. \cite{sinlapanuntakul2024}, we developed these maps and associated interface elements to actively engage participants in the collective identification of countries and campus buildings in response to our prompts within a synchronous digital environment. We outline the activities and design of each map in this section, with additional information in Procedure (section \ref{procedure}).

The world map, divided into 5 regional panels along with their associated country names representing global geography, was used for two activities:

\begin{itemize}
    \item \textbf{World Activity 1 (5 minutes)} prompted participants to discuss and identify the second most populated country in each designated region by placing star icons on those countries.
    \item \textbf{World Activity 2 (5 minutes)} presented 5 missing country names across the panels, with participants discussing and selecting from a pool of 10 options. For each panel, there were only 2 possible names, but they were not labeled. Only 5 of the provided options correctly matched the missing country names, adding an extra layer of complexity to the activity.
\end{itemize}
Similarly, the campus map, divided into 5 panels representing different parts of the campus that mirrored recognizable buildings and landscapes, was used for 2 associated activities:
\begin{itemize}
    \item \textbf{Campus Activity 1 (5 minutes)} prompted participants to discuss and identify the building with the largest square footage, including all floors, in each panel by placing star icons accordingly.
    \item \textbf{Campus Activity 2 (5 minutes)} presented 5 missing building names across the panels, with participants discussing and selecting from a pool of 10 options, half of which correctly matched the missing names, while the other half were fake names.
\end{itemize}

\begin{figure*}
  \centering
  \includegraphics[width=0.965\linewidth]{materials/maps.pdf}
  \caption{Game-Based World Map and University Campus Map on Miro.}
  \Description{The design of game-based world map and university campus map on Miro interface.}
  \label{fig:maps}
\end{figure*}

We strategically integrated motivational aspects through gamification to encourage interactive participation in our design, including a scoring system with extra incentives and various game-based mechanics \cite{sailer2013}. Each participant group was informed that, in addition to their participant compensation, they could earn an extra \$15 per person if their group achieved the highest score across all groups. This gamified approach of a scoring system and a potential monetary reward for reaching a “winning” condition introduced elements of competition and collaboration. We also carefully designed the tasks and game-based mechanics (e.g., time limit, text-based labels, the strategic use of false options, and interactive element-movement actions) to facilitate dynamic collaborative experiences, leveraging collective knowledge and engaging challenges.

\subsection{Participants}
\label{participants}

We recruited our participants through targeted advertisements on Slack and a university website supplemented by word-of-mouth recruitment. Due to the game-based map design of this study, we intentionally recruited groups of associated university students and recent graduates to ensure some familiarity with the selected campus map activity. Eligible participants signed up and participated in pre-formed groups of four, were already acquainted as friends or colleagues, and required access to Memoji on their mobile devices. During the recruitment, we informed the participants that they would be engaging in collaborative activities with their pre-formed group members to solve game-based activities. A total of 32 participants participated in this study, resulting in 8 participating groups with 4 participants per group. We present the self-reported demographics of each participant group in Table~\ref{tab:participant}.

\begin{table*}[ht]
\centering
\caption{Self-reported demographic data of participant groups. \textit{Mixed} race indicates self-identification as both Asian and White. \textit{Avatar Exp.} quantifies participants' prior experience with avatars in non-gaming video-based communication platforms.}
\Description{Table displaying self-reported demographic data, including gorup, condition order, participant ID, gender, race, and experience using avatars in non-gaming video-based platforms in hours.}
\begin{minipage}[t]{0.4925\textwidth}
    \centering
    \begin{tabular}{cccccc}
    \rowcolor[HTML]{EFEFEF} 
    \toprule
        Group & Condition Order & ID & Gender & Race & Avatar Exp. \\
    \midrule
        A & Avatar→Video & P1 & M & Asian & 10-30 hr \\
        A & Avatar→Video & P2 & M & Asian & <1-3 hr \\
        A & Avatar→Video & P3 & F & Asian & <1-3 hr \\
        A & Avatar→Video & P4 & F & Mixed & 10-30 hr \\

        \rowcolor[HTML]{EFEFEF}
        B & Video→Avatar & P5 & M & Asian & 3-5 hr \\
        \rowcolor[HTML]{EFEFEF}
        B & Video→Avatar & P6 & M & Asian & <1-3 hr \\
        \rowcolor[HTML]{EFEFEF}
        B & Video→Avatar & P7 & F & Asian & <1-3 hr \\
        \rowcolor[HTML]{EFEFEF}
        B & Video→Avatar & P8 & F & Asian & <1-3 hr \\

        C & Avatar→Video & P9 & M & Mixed & <1-3 hr \\
        C & Avatar→Video & P10 & M & Asian & <1-3 hr \\
        C & Avatar→Video & P11 & F & Mixed & N/A \\
        C & Avatar→Video & P12 & F & Mixed & <1-3 hr \\

        \rowcolor[HTML]{EFEFEF}
        D & Video→Avatar & P13 & M & Asian & <1-3 hr \\
        \rowcolor[HTML]{EFEFEF}
        D & Video→Avatar & P14 & F & Asian & N/A \\
        \rowcolor[HTML]{EFEFEF}
        D & Video→Avatar & P15 & M & Asian & 5-10 hr \\
        \rowcolor[HTML]{EFEFEF}
        D & Video→Avatar & P16 & F & Asian & N/A \\
    \bottomrule
    \end{tabular}
\end{minipage}%
\hfill%
\begin{minipage}[t]{0.4925\textwidth}
    \centering
    \begin{tabular}{cccccc}
    \rowcolor[HTML]{EFEFEF}
    \toprule
        Group & Condition Order & ID & Gender & Race & Avatar Exp. \\
    \midrule
        E & Avatar→Video & P17 & F & White & 3-5 hr \\
        E & Avatar→Video & P18 & F & White & N/A \\
        E & Avatar→Video & P19 & M & Asian & N/A \\
        E & Avatar→Video & P20 & F & Asian & N/A \\

        \rowcolor[HTML]{EFEFEF}
        F & Video→Avatar & P21 & F & White & N/A \\
        \rowcolor[HTML]{EFEFEF}
        F & Video→Avatar & P22 & M & Asian & <1-3 hr \\
        \rowcolor[HTML]{EFEFEF}
        F & Video→Avatar & P23 & F & Asian & <1-3 hr \\
        \rowcolor[HTML]{EFEFEF}
        F & Video→Avatar & P24 & M & Black & <1-3 hr \\
        
        G & Avatar→Video & P25 & F & Asian & N/A \\
        G & Avatar→Video & P26 & M & Asian & <1-3 hr \\
        G & Avatar→Video & P28 & M & Asian & <1-3 hr \\
        G & Avatar→Video & P28 & M & Asian & <1-3 hr \\
        
        \rowcolor[HTML]{EFEFEF}
        H & Video→Avatar & P29 & F & Asian & <1-3 hr \\
        \rowcolor[HTML]{EFEFEF}
        H & Video→Avatar & P30 & F & Asian & <1-3 hr \\
        \rowcolor[HTML]{EFEFEF}
        H & Video→Avatar & P31 & M & Asian & <1-3 hr \\
        \rowcolor[HTML]{EFEFEF}
        H & Video→Avatar & P32 & F & Asian & <1-3 hr \\
    \bottomrule
    \end{tabular}
\end{minipage}
\label{tab:participant}
\end{table*}

Participants identified and reported on their Memoji design in three main categories \cite{ratan2013, ratan2016}: actual self (\textit{n} = 21), ideal self (\textit{n} = 8), and creative/expressive self (\textit{n} = 3). Some also self-reported prior experience with avatar-enabled virtual conferencing platforms. This included Memoji on FaceTime (\textit{n} = 16) and avatar features on other platforms, like Google Meet (\textit{n} = 5), Microsoft Teams (\textit{n} = 1), social VR (non-gaming) platforms (\textit{n} = 6), and Zoom (\textit{n} = 7).

\subsection{Procedure}
\label{procedure}

Before the study, we instructed participants to create their own humanoid Memoji, with the flexibility to design them in ways that would make them feel comfortable and reflect their sense of self. Section \ref{participants} reports the Memoji design categories. Most of the participants (\textit{n} = 30) had already created their Memoji before being recruited to the study, with 16 reporting prior use on FaceTime. Nevertheless, we asked them to revisit and update their Memoji as desired. Data collection sessions were conducted via video-based platforms using both Zoom and FaceTime. We controlled group dynamics by recruiting groups of four and restricting the relationship of each group’s members as prior acquaintances (e.g., friends). After all group members had consented to participate, one participant per group assumed the role of \textit{operator} who manipulated elements on the game-based maps and received a brief tutorial on interacting with the Miro interface. During each session, operators were provided access to Miro containing the game-based maps and shared their Miro screen on Zoom. We conducted multiple rounds of playtesting \cite{fullerton2004} and pilot sessions on the design of the game-based maps and activities to ensure that these additional responsibilities did not alter the operator’s collaborative process and task load compared to other members. The decision-making process remained collaborative with all the group members expected to contribute equally to discussions and problem-solving regardless of the operator’s role. Participants joined the FaceTime link using their personal mobile devices, keeping their laptop cameras off and audio muted as the screens featured Miro on Zoom. Participants worked together in the same group across conditions to solve game-based activities on two maps, a world map and a university campus map, respectively, on Zoom while communicating via FaceTime. Participants' phones, displaying Memoji, were placed in front of their laptop screens, as shown in Figure \ref{fig:setup}.

During FaceTime sessions, participating groups were presented with camera-driven representation conditions (avatar vs. video), counterbalanced across all groups. The on-screen layout is represented with a 2x2 grid with each squared-grid mirroring each user’s real-time video, whether with or without a humanoid avatar representation. Both game-based maps consisted of two comparable activities each, resulting in a total of four activities (see Section \ref{map-design}). Before each map activity, participants received an explanation and prompt, with a 5-minute time limit per activity. We deliberately chose this time-constrained approach to simulate sufficient collaborative moments, as the activity's mechanism was straightforward, despite longer interactions typically occurring in real-world practice. Our focus was on participants’ perceptions during interactions, rather than activity outcomes (i.e., scores). By keeping activities concise, we maintained participation engagement throughout the study and minimized potential boredom effects.

Following the completion of each map, participants filled out a set of questionnaires, including the Rosenberg Self-Esteem Scale and the Collaboration Satisfaction. Demographic questions were included in the second set of questionnaires. Following this, we engaged participants in group interviews to share their experiences and perceptions of the collaborative interaction experience during the study group activities: one map with avatar representation and the other with video representation. Our interviews were audio-transcribed. After the group interviews, we debriefed the participants and compensated them with each receiving a \$15 gift card. Each session lasted approximately 60 minutes. The team with the highest score was awarded an additional \$15 gift card for each team member. Our study was approved by our University’s Institutional Review Board (IRB).

\subsection{Measures}

\subsubsection{Self-Esteem}

We used the Rosenberg Self-Esteem Scale \cite{rosenberg1965}, a widely validated measure of self-esteem, to measure participants’ positive and negative feelings about self-confidence and self-image after engaging in each of the maps. The original 10-item scale, rated on a 4-point Guttman-type rating scale (strongly agree to strongly disagree), was adjusted by adding the phrase “during the recently completed group activity” to each item, capturing a more contextually relevant assessment of participants’ self-reflection within collaborative experiences of each group activity. The total sum of possible score on the scale was 40, with higher scores indicating greater self-esteem.

\subsubsection{Video-Based Collaboration Satisfaction}

The Collaboration Satisfaction instrument \cite{reinig2003} measured collaboration satisfaction with the process and outcomes of group work comprising 8 items on a 5-point scale that rate two constructs: satisfaction with collaborative process (SP) and satisfaction with collaborative decision (SD). SP construct includes 5 semantic differential items representing group dynamics during collaborative interactions, while SD includes 3 refined Likert scale items (from an originally intended 5, based on poor loading in factor analysis \cite{green1980}) assessing satisfaction with outcomes. We adjusted items for context-specific relevance: SP references “the recently completed group activity,” and SD replaces “group’s solution” with “group’s decision.” These adjustments aimed to more accurately capture participants’ perceptions. The total score is calculated as an average across all items, with higher average scores indicating a greater level of satisfaction with the group collaboration.

\subsubsection{Group Interview}
\label{group-interview}

Our group interviews were meticulously structured around a set of questions designed to explore participants’ experiences and perceptions of avatar-mediated communication within collaborative group settings. The main interview questions are outlined in Appendix \ref{appendix:group-interview}. To ensure rigor and reliability, we iteratively designed, piloted, and refined the interview questions to ensure they were clear, framing them in a way that would elicit neutral, unbiased responses.

The question set was consistent across all groups, regardless of their background, with tailored follow-up questions to align with discussion topics during the interviews. We chose group interviews over individual ones to reflect the collective experience and collaborative nature of study activities, stimulating richer discussions by leveraging pre-existing rapport within their groups given that they were already acquainted. This format enabled more candid, contextualized insights and immediate comparison of perspectives on avatar use in established teams, despite potential group dynamics' influence on individual responses.

\subsection{Data Analysis}

\subsubsection{Quantitative Analysis}

We conducted our primary analyses using paired-samples t-tests at the individual level to assess within-subject differences in self-esteem and video-based collaboration satisfaction between the study conditions (avatar vs. traditional video). This analytical approach aligns with our research questions, which focus specifically on individual experiences and perceptions within collaborative interactions rather than group-level dynamics. Prior to analysis, we conducted Shapiro-Wilk tests to check normality distributions (Table \ref{tab:normality}) with \textit{p} > .05 across all measures, confirming that parametric tests were appropriate for our data. 

\begin{table}[H]
\centering
\caption{Shapiro-Wilk tests for paired differences. Note: \textit{p} > .05 indicates that the data satisfy normal assumptions.}
\Description{Table displaying Shapiro-Wilk test results with \textit{p}-values.}
\label{tab:normality}
\begin{tabular}{lcc}
\rowcolor[HTML]{EFEFEF} 
\toprule
Measure & Shapiro-Wilk W & \textit{p}-value \\
\midrule
Self-Esteem & 0.963 & .328 \\
Collaboration Satisfaction & 0.952 & .167 \\
Satisfaction w/ Collab. Process & 0.959 & .261 \\
Satisfaction w/ Collab. Decision & 0.948 & .128 \\
\bottomrule
\end{tabular}
\end{table}

While our research questions focus on individual experiences, we calculated intraclass correlation coefficients (ICCs) to assess the degree of non-independence within groups. The ICC values (self-esteem: 0.611; overall satisfaction: 0.720; process satisfaction: 0.754; decision satisfaction: 0.677) indicated substantial group-level dependency, as would be expected in social settings. However, given our research questions specifically target individual experiences in social contexts rather than group-level phenomena, and the limited number of groups (\textit{N} = 8) in our study, our primary analyses utilize individual-level paired t-tests, with the acknowledgment that this approach may not fully account for group-level variance.
A priori power analysis indicated that our sample size (\textit{N} = 32 participants) was sufficient for detecting medium-to-large effects (\textit{d} $\geq$ 0.60) at $\alpha$ = .05 with power of .80 for paired t-tests at the individual level. Cohen's \textit{d} \cite{cohen1988} was calculated to quantify the effect size of the observed mean differences. For correlational analyses, we used Pearson correlation (\textit{r}) with Bonferroni corrections to adjust for the family-wise error rate.

\subsubsection{Qualitative Analysis}

We conducted a combined approach of inductive summary annotation coding and deductive reflexive thematic analysis \cite{braun2006} to explore participants’ perceptions and experiences with avatars in collaborative videoconferencing interactions. Given that the first author solely conducted all group interviews, our approach prioritized reflexivity and depth in concept and theme generation to allow for nuanced interpretations of each code instead of pursuing inter-rater reliability \cite{mcdonald2019}. 

We followed a structured procedure for thematic analysis, beginning with the first author reviewing all eight group interview transcripts and segmenting them according to the session's structured and follow-up questions (see Appendix~\ref{appendix:group-interview} for the interview protocol). In the second phase, 20 student researchers performed independent coding using summary annotations. To ensure robustness and mitigate individual bias, each transcript was coded five times, with student researchers randomly assigned two transcripts, yielding 40 coded instances. Although most student researchers had prior qualitative coding knowledge and experience, all participated in a 60-minute training session, which guided them to carefully analyzing each data segment. Following this, they independently developed analytical memos \cite{birks2008}, annotating summaries and any points of divergence for further interpretation. Subsequently, our third step involved the first author reviewing these annotations to identify recurrent themes. Following this, the first author revisited the original eight group interview transcripts to refine thematic topics and develop sub-themes, subsequently engaging in discussions with the student researchers to iterate on these initial themes. Lastly, both authors collaboratively reviewed and further refined the themes, descriptions, and interpretations, concerning the experiences of using avatars in video-based collaborative interactions.

\section{Results}

\subsection{Impact of Avatars on Self-Esteem and Collaboration Satisfaction}

Responding to RQ1, our findings at the individual level revealed significantly higher reported scores in both self-esteem and collaboration satisfaction when using avatar than when using traditional video representation. Analysis of condition order effects revealed no significant differences between avatar-first and video-first groups for self-esteem (\textit{t}(30) = 1.11, \textit{p} = .274) or collaboration satisfaction (\textit{t}(30) = 1.33, \textit{p} = .192), confirming our counterbalancing strategy effectively controlled for sequence effects.

\subsubsection{Self-Esteem}

\label{self-esteem}

Paired-samples t-test revealed a statistically significant difference in self-esteem scores between the avatar (\textit{M} = 31.84, \textit{SD} = 4.55) and the traditional video representation conditions (\textit{M} = 26.97, \textit{SD} = 4.72), \textit{t}(31) = 4.67, \textit{p} < .001; \textit{d} = 1.05, with a large effect size (Figure \ref{fig:self-esteem}). The ICC for self-esteem (0.611) indicated substantial group-level dependency, consistent with the social nature of the experimental context. This finding demonstrates that participants experienced higher levels of self-esteem and self-confidence when visually represented with avatars compared to live video of themselves during video-based collaboration, potentially influencing their engagement and experiences in collaborative activities.

Although not significantly different, we observed a trend in which “operators” rated their self-esteem higher than other group members in both study conditions. In the avatar condition, operators’ self-esteem ratings (\textit{M} = 33.88, \textit{SD} = 4.79) exceeded those of other members (\textit{M} = 31.17, \textit{SD} = 4.36). A similar pattern appeared in the video condition (operators: M = 29.50, \textit{SD} = 4.31; other group members: \textit{M} = 26.13, \textit{SD} = 4.63). This trend may warrant further investigation, potentially stemming from operators’ perceived greater control and responsibility within the group.

\begin{figure}[H]
  \centering
  \includegraphics[width=\linewidth]{materials/self-esteem.pdf}
  \caption{Comparison of average self-esteem ratings. Error bars represent +/- 1 standard error.}
  \Description{Bar graphs comparing self-esteem ratings between avatar and video representations. Avatar representation demonstrates significantly higher self-esteem ratings compared to video representation, p < .001.}
  \label{fig:self-esteem}
\end{figure}

\subsubsection{Collaboration Satisfaction}

\label{satisfaction}

At the individual level, we found significant differences in video-based collaboration satisfaction between camera-driven representation conditions. With the highest and lowest possible scores of 5 and 1, respectively, participants reported significantly higher overall collaboration satisfaction with avatar representation (\textit{M} = 4.13, \textit{SD} = 0.61) compared to video representation (\textit{M} = 3.66, \textit{SD} = 0.66), \textit{t}(31) = 2.81, \textit{p} = .008; \textit{d} = 0.74 (Figure \ref{fig:satisfaction}). The ICC for this measure (0.720) indicated substantial group-level dependency, reflecting the shared contextual factors inherent in collaborative interactions. This finding suggests that the choice of camera-driven representation has a discernible impact on individual satisfaction in video-based collaboration, with avatar representation significantly enhancing overall collaboration satisfaction among participants.

\begin{figure*}
  \centering
  \includegraphics[width=0.9\linewidth]{materials/satisfaction.pdf}
  \caption{Comparison of average video-based collaboration satisfaction ratings. Error bars represent +/- 1 standard error.}
  \Description{Bar graphs comparing avatar and video representations in terms of satisfaction with overall collaboration, collaborative process, and collaborative decision-making. Compared to video representation, avatar representation demonstrates significantly higher ratings in satisfaction with overall collaboration (p = .008), with collaborative process (p = .012), and with collaborative decision (p = .014).}
  \label{fig:satisfaction}
\end{figure*}

Our analyses of specific aspects of video-based collaboration satisfaction revealed similar patterns (Figure \ref{fig:satisfaction}). The satisfaction with collaborative process was significantly higher in the avatar representation (\textit{M} = 4.25, \textit{SD} = 0.60) compared to the traditional video condition (\textit{M} = 3.81, \textit{SD} = 0.66), \textit{t}(31) = 2.68, \textit{p} = .012; \textit{d} = 0.71. The ICC for this measure (0.754) was high, indicating strong within-group consistency in process evaluations. This finding suggests that avatars positively influence individual perceptions of the video-based collaborative process, which include perceived efficiency, coordination, fairness, understanding, and satisfaction.

Similarly, the satisfaction with collaborative decisions in the avatar condition (\textit{M} = 3.94, \textit{SD} = 0.73) was significantly higher than the video condition (\textit{M} = 3.43, \textit{SD} = 0.88), \textit{t}(31) = 2.61, \textit{p} = .014; \textit{d} = 0.63, indicating a medium effect size. The ICC (0.677) indicated substantial group-level dependency, as expected for shared decision processes. This finding suggests that avatar representation contributes to higher individual decision-making satisfaction during video-based collaborations. This could be related to the extent to which participants feel their inputs are reflected in the final decision, their sense of commitment to those decisions, and their belief and trust in the correctness of the group’s choices made.

\subsubsection{Interrelationships between Self-Esteem and Collaboration Satisfaction}

\label{interrelationships between self-esteem and satisfaction}

We used Pearson correlation coefficients to assess the relationships between self-esteem, collaboration satisfaction, and decision-making processes in both avatar and video representation conditions. Particularly noteworthy were the strong positive correlations observed in the avatar representation between self-esteem and dimensions of video-based collaboration satisfaction: overall satisfaction, \textit{r}(30) = .72, \textit{p} < .001; process satisfaction, \textit{r}(30) = .72, \textit{p} < .001; and decision-making satisfaction, \textit{r}(30) = .62, \textit{p} < .001. Our findings suggest that, in the context of avatar representation, higher levels of perceived self-esteem align with increased satisfaction in group collaboration, process, and decision-making experiences at the individual level. However, several other correlations either failed to reach statistical significance after Bonferroni correction or exhibited weak correlations in the video representation condition, implying that this relationship may be unique to the avatar-mediated environment.

The distinction in the mechanisms driving higher self-esteem and increased collaboration satisfaction suggests that these two outcomes may be influenced by different factors, providing an explanation for their lack of correlation as demonstrated. The elevated self-esteem might be attributed to the overall cute-looking visual of the avatars, aligning with the Proteus effect’s proposition that physical characteristics and anticipated behaviors can influence the behavior and cognition of individuals \cite{liu2023, praetorius2021, yee2007}. 

In contrast, we suspect that the enhanced collaboration satisfaction could be associated with a degree of visual consistency among avatars across participants, creating a higher sense of group or shared identity. This aligns with social identity theory \cite{ashforth1989}, which posits that similarity in representations contributes to the formation of a shared social identity and a sense of belonging within a collaborative setting. Participants had a pre-existing relationship within their groups prior to the study, yet avatars appear to have reduced social comparison within groups. This likely redirected attention toward the content of communication and the collaborative process, ultimately increasing individual satisfaction with the interaction.

\subsection{Benefits in Avatar-Mediated Collaborative Interactions}

\label{benefits}

In addition to the measured effects of avatars (detailed in Sections \ref{self-esteem}, \ref{satisfaction}, and \ref{interrelationships between self-esteem and satisfaction}), we further explored dimensions of the experience via group interviews. Here, we address a part of RQ2 by discussing observed themes on avatars as mediators to enhance self-awareness, engagement, and comfort while alleviating social awkwardness.

\subsubsection{Increased Self-Awareness and Confidence}

An unexpected outcome was the dynamic nature of self-awareness and confidence reported by participants while using avatars, which aligns with objective self-awareness theory \cite{duvalandwicklund1972}. 
Participants initially experienced heightened public self-awareness, becoming conscious of their actions as objects in the external world, often through the perceptions of others: “\textit{I think I paid more attention to my own expressions. I didn’t wanna come across like I was frowning, and avatars sort of forced me to, maybe, smile a little more. So, it'd show up}” (P8). This indicates participants' conscious effort to manage a positive demeanor, likely driven by an acute awareness of how their avatar might be perceived by others.

As participants became more accustomed to the avatars, their focus shifted from public self-awareness to a more natural state of private self-awareness, referring to the internal awareness of their own thoughts, feelings, and bodily states. This transition often led to increased self-confidence and more authentic self-expression. P22 explained, “\textit{it is like I’m having a mask on, and so in a way I’m more confident with my own thoughts. Like the confidence to say it out loud}.” This metaphor suggests that, as they became more comfortable with their avatar representations, they felt empowered to express themselves more openly, suggesting that this newfound confidence allowed them to reveal a more authentic self.

A few others discussed a shift in attention from the game-based task to group dynamics. P32 observed, “\textit{I wasn’t a hundred percent focused on the task. I would go back to looking at it and at other people’s, but I feel like I paid more attention to what others are doing},” suggesting an increased awareness of group interactions mediated by avatars. Interestingly, this heightened self-awareness often manifested as a form of self-prioritization, where participants' attention was predominantly focused on their own avatar representation and behavior rather than on others or the collaborative task, despite being in a familiar group setting. This progression from heightened public self-awareness to a more comfortable state of private self-awareness indicates that the observed effects are not inherent properties of the Memojis, but rather reflect early impressions of using them as participants adapt to this form of representation.

\subsubsection{From Self-Consciousness to Engaged Participation}

The initial self-consciousness experienced by participants in response to avatars evolved over time, aligning with the transition from public to private self-awareness and the objective self-awareness theory \cite{duvalandwicklund1972}. Early on, participants were preoccupied with their avatars, actively experimenting with positioning and real-time facial tracking, as seen when one participant expressed concern over their avatar being partially or completely out of the frame on their device screen: “\textit{I was more conscious because I was leaning forward [...] like, am I out of frame on my phone? The Memoji wasn’t on my face all the time.}” (P31).

This increased self-monitoring initially led to performance anxiety as participants focused on their avatar representation \cite{duvalandwicklund1972}. However, as they became more familiar with the technology, their focus shifted from self-presentation to communication, and self-consciousness diminished with participants “\textit{worrying about it less}” (P19, P20), evolving into a more casual aspect of their digitized interactions. This transition to private self-awareness enabled more meaningful interactions, reducing self-monitoring, as participants had “\textit{more time to think through the [collaborative] activities}” (P18). This progression from heightened public self-awareness to a more comfortable state of private self-awareness indicates that the observed effects are not inherent properties of the Memojis, but rather reflect early impressions of using them. Participants likely became more adept at controlling their avatar’s expressions as they gained more experience using their Memojis.

\subsubsection{Comfort in Augmented Appearance}

With the rise of video conferencing platforms, our physical appearance in video-based interactions has become a central concern. For some, this shift has brought about a sense of unease and self-consciousness, connecting one’s authentic self and the pressures of presenting to larger audiences. The act and need of turning on a camera to work with others can be an anxiety-inducing experience, as individuals may feel anxious of being judged based on their physical attributes or actions. As P28 noted, avatars could serve as a solution to this problem: 

\begin{quote}
    “When we turn our camera on, we actually pay a lot of attention to our features [...] I think avatars will be very useful when someone does not really want to show their face when in a spotlight” (P28).
\end{quote}

P11 took this notion further, speculating that, “\textit{[an avatar] could be useful [...] if someone does not really feel confident with their appearance}.” This augmentation empowers individuals to present themselves in a manner that aligns with their self-image, thus mitigating the burden of self-doubt. Some participants, drawing on both past experiences and their participation in this study, pointed out how using avatars could mitigate self-consciousness when having a camera on during video-based work interactions. For instance:

\begin{quote}
    “If my camera is on during video meetings for work, I get conscious sometimes and I’m like, what is my face doing? So, I thought the avatar actually provided that mask where the animation is not gonna look bad no matter what angle it is” (P29).
\end{quote}

\subsubsection{Reduced Social Tension and Awkwardness}

In video-based collaboration, avatars were viewed as an effective strategy for mitigating social tension and alleviating awkwardness. Participants noted that these avatars had the ability to “\textit{create a relaxed atmosphere}” (P10, P24) during interactions, helping to enhance group dynamics and reduce anxiety often associated with such collaboration. Participants viewed these avatars as instrumental in steering conversations towards more engaging and diverse topics, encouraging interactions beyond typical work-related subjects. They believed that recurring team avatars could act as “\textit{ice breakers, particularly in situations where participants were less familiar with each other}” (P30), essentially “\textit{stimulating creativity and promoting camaraderie within the group}” (P9). The influence of avatars extended beyond being mere “ice breaker,” with several of our participants reflecting that avatars made the collaboration experience “\textit{feel less like work and more like to a group bonding session}” (P1).

Participants generally did not perceive avatars as an essential component throughout the activities but as “\textit{a valuable tool during specific segments}” (P27). They suggested that integrating avatars in the initial minutes of collaborative activities could set a positive tone, invoke laughter, and elevate the overall atmosphere. They viewed avatars as particularly effective in interactions designed for team-building in activities, such as “\textit{fun Fridays}” (P15) or “\textit{themed meetings}” (P14). Furthermore, avatars were regarded as “\textit{a means of stress reduction}” (P2). The inherent charm of these whimsical representations alleviated tension and anxiety, even when the mediated expressions were intended to be serious (P3).

\subsection{Challenges in Avatar-Mediated Expression and Interpretation}

\label{challenges}

Despite the benefits that avatars offer, participants acknowledged the challenges associated with avatar-mediated communication. In this section, we continue to respond to RQ2 by discussing the critical aspects of avatar representation in both expressing and perceiving emotions to maintain effective collaborative interactions.

\subsubsection{Inaccuracies in Conveying Expression}

Avatars have emerged as important tools for conveying facial expressions within video-based meetings, yet challenges surfaced in participants' attempts to accurately express emotions, contributing to occasional distracting self-prioritization. The difficulty in conveying nuanced emotional states underscore technological limitations in achieving seamless communication. P31 astutely noted: “\textit{Your avatar brings more attention to certain expressions you’re making. It shows a natural smile on your face when your face is just resting. [It] kind of accentuates that}.” While the emotive realism that avatars bring to mediated interactions fails to accurately capture subtle emotional nuances, it may overamplify certain expressions. Similarly, P8 pointed out that avatars sometimes misrepresented facial expressions: “\textit{I saw my avatar frowning when, in fact, I just had a natural, resting feel on. I had to put up a smile so it’d translate onto the screen}” (P8).

In contrast, while tone and expressions are fundamental for understanding emotional states, some participants raised a pertinent concern that avatars, correctly reflecting head and mouth movements, often neglect the expressive power of the eyes. For instance, P25 mentioned, “\textit{avatars don’t usually convey too much expression, except for the head movements and mouth, and we express emotions with other parts that don’t get captured, for example, the eyes}.” There are mixed viewpoints on avatar fidelity in emotional expression. While avatars hold the potential to enhance emotional expression and facilitate communication, they bear technological limitations in capturing the full complexity of human emotion.

\subsubsection{Difficulty in Interpreting Expression}

The immersive experience of avatars posed challenges in accurately deciphering the thoughts and emotions of others. Participants expressed the need for more cues to improve the interpretability of non-verbal communication. Interestingly, P19 observed through trial and error that reading emotions through avatars was “\textit{quite similar to reading faces [...] but required less cognitive effort}.” Despite this, several participants found the large Memoji heads “\textit{easier to see who’s listening from nodding}” (P3, P4), while some struggled with the limited ability to express subtle non-verbal cues. For example, P26 mentioned: 

\begin{quote}
    “The only difference would be the subtle nods that people make when they’re talking, which we couldn’t really see through the Memoji. It was hard. You had to even say that you were still thinking” (P26). 
\end{quote}
This led to many relying more on audio in both delivering and receiving messages. Moreover, participants mentioned the importance of accurately perceiving the emotions of others in collaborative activities, noting that avatars obscured true feeling. One participant reflected:

\begin{quote}
    “It is hard to gauge the person’s true feelings about any particular thing. Because it’s very subtle, I think the changes in expressions are barely noticeable without paying close attention. In a collaborative activity, I think it is important to know everyone’s true thoughts because that can help you arrive at a more accurate decision” (P25).
\end{quote}

Participants experienced difficulties in interpreting emotions through avatars. For instance, P22 stated, “\textit{I find it harder to see their expressions. Sometimes, I’m unsure if they’re thinking or if the Memoji simply isn’t reacting}.” Similarly, P16 echoes this: “\textit{I sense that the pre-scripted remote responses often transform a small smile into the cheesiest grin you’ve ever seen. It all feels somewhat off, and I couldn’t read people as well}.” The ability to decode subtle facial expressions in group settings is important for collaborative interactions because “\textit{we express our agreements, disagreements, and hunches with our faces}” (P24). Without seeing actual faces, comprehending others’ thoughts becomes more challenging. Likewise, P21 emphasized the importance of visual cues in time-sensitive activities: 

\begin{quote}
    “It was harder to tell how other people were feeling [...] Being able to see people’s faces is really important, especially if you are working on something that’s timed; you need all the information there in front of you” (P21).
\end{quote}

\subsubsection{Fidelity and Authenticity}

Avatar fidelity becomes a pivotal factor in creating truly authentic emotional expression. Participants identified an inherent challenge in ensuring avatars accurately reflected their emotions and identities, introducing tension in video-based collaborative interactions. P13 observed, “\textit{it feels weird when the Memoji don’t really look exactly like the person [...] it just doesn’t feel as authentic as talking face to face. It almost feels like they’re not really there [but] more like their voice coming out of an avatar}.” When technology attempts to replicate human identity, even slight deviations from authenticity can disrupt the sense of genuine presence. Perceiving such deviations–even subtly–can produce an underlying uneasiness, a feeling that the other person on the screen lacks true presence, noting that authentic human presence does not only involve faithfully replicating one’s appearance. Instead, P20 suggested that embracing a whimsical and unconventional avatar can sometimes promote a more authentic sense of self-expression in an absurd way:

\begin{quote}
    “From my past experience with avatars, I’m more likely to use the animal-like ones because I feel like they really emphasize the overall goofiness of me. I feel like using the one that looks exactly like me feels a little off because like I’m right here. But if I could look like a giraffe, that’s kind of fun” (P20).
\end{quote}

Such thinking indicates that embodying avatars differing from one's actual self may allow individuals to drop their guards and encourage reveal their genuine selves. Authenticity in mediated interactions does not always require literal representation; rather, it emerges from the freedom to be unfiltered in order to create profound and meaningful connections.

\subsection{Influential Factors for Avatar Adoption}

\label{factors}

In terms of the factors influencing the likelihood to use avatars in video-based meetings, aligned with RQ3, we discovered several key insights that provide a nuanced understanding of the avatar adoption. We asked participants whether they would be likely to use an avatar while video conferencing if (1) others were also using avatars, (2) the avatar could be easily customized before joining a video conference, and (3) the avatar represented facial expressions with greater accuracy, respectively.

\subsubsection{Social Influence, Customization, and Accuracy}

Social influence emerged as a pivotal determinant of avatar adoption, with unanimous agreement from all 32 participants (100\%) that they would be more likely to use an avatar if others were also using them. The consensus highlights the fundamentally social nature and peer influence in shaping an individual’s inclination to use avatars within collaborative settings.

With 31 out of 32 participants (97\%) expressing agreement that they would be more likely to use an avatar if they could easily customize it before joining a video conference, the ability to personalize and tailor avatars to align with individual preferences and identities was strongly valued. One participant, however, thought otherwise and disagreed because they “\textit{would prefer to just not use an avatar at all}” (P31).

Across all participant groups, 28 out of 32 participants (88\%) agreed they would more likely use an avatar if it accurately represented their facial expressions, acknowledging accuracy in emotion expression as a critical determinant for avatar usage. Participant P7 aptly expressed this sentiment: “\textit{If it could pick up the exact level of expression, that would be ideal}.” The ability of avatars to precisely and authentically mirror micro-level emotions proved instrumental in conveying the subtleties of non-verbal communication. This factor significantly shaped users’ perceptions of avatars’ efficacy in creating genuine human connections within video meetings.

However, one group (\textit{n} = 4) held a dissenting opinion and argued that human interaction in its natural form, unaltered by avatars, was the superior medium for intended emotional conveyance. P29, for instance, expressed a preference to use traditional video interfaces instead of an avatar with greater emotion expression accuracy: “\textit{I disagree. If I am intending on having a real conversation, I would probably just want my face}” (P29).

\subsubsection{Contextual Appropriateness}

Participants imagined possible scenarios to speculate about the appropriateness of using avatars in video-mediated environments, considering their suitability in different contexts. It was evident that the acceptability of avatars depended on the nature of the interaction. Participants acknowledged that avatars were more aligned with casual or informal settings, but the extent to which they were deemed suitable in professional contexts varied.

In work organizations with strong traditions of formality, avatars may be perceived as less appropriate due to their often cartoonish nature, which can be seen as incongruous with professional settings \cite{junuzovic2012}. Some participants questioned the appropriateness of avatar use in such environments because it might appear “\textit{silly to discuss serious topics, such as financial or healthcare matters}” (P3). In addition, avatars are deemed as better suited for light-hearted interactions or as a means to infuse fun into conversations within a familiar context. Participants noted that avatars are “\textit{perfect for a good night out with friends in FaceTime events}” (P22) and that they would use avatars with those whom they feel comfortable, such as “\textit{sisters, mom, and dad, as they feel more casual than formal}” (P24).

Some participants aptly highlighted the pivotal role of organizational norms and practices on the acceptability of avatars in video-based collaborations. As P2 suggested, “\textit{if the team is more traditional or formal, using avatars might not be appropriate. But using avatars in the team that values creativity and innovation could be fun and engaging}.” Moreover, participants noted that the use of avatars in more formal work-related situations is viewed with skepticism, as the presence of avatars can make it challenging to convey and interpret serious or critical information. P6 imagined interacting with a manager who uses an avatar representation while assigning work and providing instructions:

\begin{quote}
    “[The avatar] might be shouting at you. But with a very green hair mohawk-style Memoji, which would obviously be funny, I wouldn’t be able to take the critique. No matter how well it replicates their real expressions, like anger or frowning, I would still think it’s adorable” (P6).
\end{quote}
The appropriateness of using avatars in video meetings is contingent on factors such as the nature of the interaction. While there was consensus that avatars are well-suited for casual or informal settings, the tension between the perceived unseriousness and the potential for enhancing engagement and creativity contribute to the evolving nature of video-based communication norms, where context plays a key role in shaping the acceptance of avatars in collaborative interactions.

\section{Discussion}

Our study reveals the significant impact of avatar representation over traditional video representation on self-esteem and collaboration satisfaction in video-based meetings, while also identifying the associated benefits, challenges, and influencing factors. This section discusses: (1) key findings, reflections, and their connections with existing research; (2) avatars as an alternative solution between camera on and off; and (3) the unexpected emergence of a self-prioritization in avatar-mediated collaborative interactions.

\subsection{Avatars Effects on Self-Perception and Collaboration}

Unlike prior research \cite{bente2008, inkpen2011, junuzovic2012} comparing different user representation modalities, namely VR avatars, video, and audio, our work uniquely investigates how interacting via avatars, compared to traditional video interfaces, affects self-perception and collaboration experiences in non-immersive video-based meetings.

Our findings demonstrate significant positive impacts of avatars on self-esteem and collaboration satisfaction at the individual level, with large effect sizes. Elevated self-esteem suggests that avatar representation positively shapes engagement and experiences in collaborative activities. This enhanced self-perception supports self-discrepancy theory \cite{higgins1987}, which posits that avatars enable users to project representations closer to their ideal selves, thereby reducing discrepancies between actual and desired self-images and effectively boosting self-esteem in video-based interactions. We recommend further research to look into how avatar customization and control enhance feelings of empowerment and self-efficacy \cite{bandura1997}, which in turn influence self-esteem. In addition, avatars significantly enhanced multiple dimensions of collaboration satisfaction, including efficiency, coordination, fairness, understanding, and contentment in group decisions, greater commitment to those decisions, and higher levels of trust in their correctness.

Our findings suggest avatars are perceived as valuable in informal, social, and low-stakes contexts where rapport and comfort are prioritized over professional presentation. While avatar representation is preferred when visual representation is otherwise lacking \cite{panda2022}, we recommend teams to carefully consider their implementation context and purpose, as avatars may be most appropriate for team-building, brainstorming sessions, and social gatherings rather than in formal, work contexts.

\subsubsection{Reflection on Temporal Progression: Reconciling Benefits and Challenges}

What might appear to be a contradiction between higher self-esteem measurements and reported difficulties with expression and self-consciousness is not a negating contradiction but rather a reflection of the complex and multifaceted nature of avatar-mediated interaction. We interpret these findings not as inconsistent, but as complementary perspectives that, together, provide a more complete understanding of how avatars simultaneously reduce appearance-based anxiety while introducing new challenges in emotional/expression authenticity. This apparent contradiction can be further reconciled through the lens of temporal development. Participants described a progression from initial heightened public self-awareness, causing self-consciousness, to a more comfortable state of private self-awareness that ultimately led to increased self-esteem. Rather than contradicting each other, the qualitative findings provide nuanced contexts of how initial adaptation challenges give way to measurable benefits over time.

Moreover, participants could simultaneously experience both benefits (e.g., enhanced self-esteem, reduced social tensions) and limitations (e.g., expression difficulties) while still reporting overall higher satisfaction with the avatar condition. 
This multidimensional experience explains why participants might verbally express concerns about certain aspects of avatar use, with concerns inherent in technology-mediated social interaction, while still rating their overall experience more positively than traditional video interactions, They continuously integrate and reconcile multiple experiential states, which may account for this apparent tension.

\subsubsection{Theoretical Mechanisms of Avatar-Mediated Self-Perception}

The temporal progression we observed aligns with and extends objective self-awareness theory \cite{duvalandwicklund1972}, which delineates crucial distinctions between public self-awareness (the conscious perception of oneself as a social object evaluated by others) and private self-awareness (introspective awareness of one's internal states, thoughts, and emotions). The transition between these metacognitive states constitutes a critical psychological mechanism underlying the benefits of avatars. Our findings extend this theoretical framework by demonstrating that in avatar-mediated contexts, this transition occurs naturally as users integrate their digital representation into their self-concept, moving from externally-oriented self-monitoring to more authentic, internally-guided engagement.

Our results also support and extend the Proteus effect \cite{liu2023, praetorius2021, yee2007}, which posits that individuals internalize and behaviorally conform to the perceived characteristics and social expectations associated with their digital representations. Whereas previous studies primarily examined this phenomenon in immersive VR environments \cite{dunn2012, fokides2021, fox2013, kocur2020, lemenager2020, messinger2008, nowak2018, praetorius2020, ratan2014, sibilla2018}, our study demonstrates that the Proteus effect also manifests through avatars within non-immersive video-mediated interactions, thereby bridging the gap between fully immersive VR experiences and non-immersive 2D interfaces.

Our findings further support previous studies \cite{park2023, ratan2022, wicks2021} on avatars' potential to mitigate anxiety in video-based meetings but also extend this understanding to group settings, where interaction dynamics differ from dyadic encounters \cite{park2023}. Avatars provided not only a sense of comfort but also facilitated positive social dynamics, with participants reporting decreased self-consciousness and enhanced interpersonal ease, which are particularly relevant for addressing the widely-documented phenomenon of Zoom anxiety \cite{falzone2005, ramasubramanian2020} that has emerged as a significant concern in remote work contexts.

Despite these benefits, we acknowledge that technological constraints at the time of data collection posed challenges in accurately conveying expressions through avatars. We recommend that designers of avatar technologies prioritize emotional expression features to address the gap in emotional nuance identified in previous studies \cite{fribourg2021, gonzalez2020facial}. Our findings indicate that, while avatars can improve self-esteem and reduce anxiety, further advancements are needed to fully capture the complexity of emotional communication. For researchers, these mixed outcomes highlight the complexities of avatar-mediated communication, where the advantages of increased self-esteem and comfort must be weighed against the limitations in emotional conveyance. We recommend and call for collaborative efforts between designers and researchers in identifying the optimal and appropriate balance.

\subsection{Avatars as a Middle Ground Between Camera On and Off}

The camera on/off dilemma represents a central challenge in real-time video-based interactions, particularly with the rise of video conferencing. As revealed in our study, researchers should consider avatars as a compelling solution to this issue, offering an alternative between full video and camera-off modes. Balancing between self-consciousness and the necessity of being \textit{presentable}, offering visual cues, remains a considerable challenge. On a similar note, research on Zoom fatigue \cite{bailenson2021, ratan2022, wicks2021} have proposed avatars as a potential replacement for the camera-off modality, supporting previous work comparing different camera-driven representations \cite{inkpen2011, junuzovic2012, panda2022}. 

Avatars challenge this binary conception of presence and absence in mediated communication by introducing a third modality of \textit{filtered presence}. Unlike the \textit{camera-off mode} that eliminates visual social cues entirely or the \textit{camera-on mode} that may induce discomfort, avatar representation allows for selective transmission of certain communicative elements, while obscuring others. It reflects behavioral nuances and maintains focus through subtle movements, enhancing engagement without requiring full camera visibility. he resulting increase in self-esteem and collaboration satisfaction observed in the avatar condition suggests that this filtered presence creates a psychologically safer space for interaction.

Designers of video-based interactive systems could use this finding to incorporate avatars as a flexible communication tool that balances the binary choices of having the camera either on or off. For example, one participant echoed this sentiment, speaking up about the concept of avatars as an effective middle ground:

\begin{quote}
    “Since working remotely has become a thing, a lot of people tend to keep their cameras off. I think avatars could be a good alternative. Rather than having no access to somebody’s facial expressions, they could bring in that element to some extent” (P9).
\end{quote}
The insight emphasizes avatars as a tool for enhanced social connection in remote environments that balances privacy concerns with communicative needs. While avatars cannot fully replicate human expressions, they offer a level of visual interaction absent in static images or blank screens \cite{panda2022}. Animated facial expressions and gestures provide a visual link that signifies presence, reducing the need for constant camera engagement. Avatars also offer greater privacy control, potentially alleviating the self-consciousness and fatigue often associated with traditional video conferencing \cite{cho2023, hooi2014, leong2023, maloney2020, ngien2023, ratan2022}.

Researchers should carefully assess the applicability of avatars for different contexts, taking into account factors such as organizational norms, team dynamics, and individual preferences. The acknowledgment of avatars as a middle ground in video-based collaboration speaks to its evolving nature and the growing need for flexibility and personalization in communication tools. By bridging the divide between active camera and camera-off modes, avatars empower individual agency and inclusivity in shaping their virtual collaborative experience.

\subsection{Self-Prioritization in Collaborative Interactions}

Our study revealed a discernible shift of self-prioritization in avatar-mediated collaborative interactions, where participants exhibited increased focus on their own avatar's expressions and appearance, often at the expense of focusing on others. This unexpected behavior, emerged even in familiar group settings with shared, goal-directed activities, challenging prior research that positioned avatars as tools for enhancing shared attention and group engagement by diverting focus from the self \cite{ferrer2022, ratan2022, wicks2021}. Researchers should explore how the advent of advanced avatars, capable of accurate facial expression tracking without obscuring the physical surroundings, redefines social dynamics and re-evaluates design strategies to balance individual self-awareness with collective interaction goals.

Notably, this self-prioritization occurred in small groups with pre-established relationships (i.e., friends or colleagues), where collective focus facilitated by pre-existing social connections and familiarity was anticipated to dominate \cite{sainsbury2009}. Emotional feedback mechanisms mediated by avatars, as indicated in \cite{benke2020, kim2019, liu2018}, contributed to heightened self-awareness, particularly with regard to inadvertent concerns about negative emotion leakage. Designers of avatar technologies could leverage this plausible explanation to ensure that such mechanisms are attuned to reduce apprehension and promote a more natural flow of interaction. Moreover, our findings add an additional layer of complexity in avatar-mediated interactions. In line with the Proteus effect \cite{yee2007}, employing avatars with cute-looking designs may inadvertently place a burden on users to express positive or \textit{happy} emotions, aligned with the avatar’s persona. This, thereby, introduces a potential coercion process where user agency is subtly cajoled by the mediator, reinforcing self-prioritization observed in this study. 

Designers and researchers alike should build on these insights to rethink how avatar-mediated systems are developed for collaborative settings. Specifically, researchers should explore how subtle nudges in avatars influence group dynamics and emotional authenticity over time. Meanwhile, designers of video-based collaborative systems should carefully balance mechanisms that encourage positivity with triggers that safeguard user agency and authentic emotional expression.

\section{Limitations and Future Research}

Our work presents new insights into the impact of avatar representation on video-mediated collaborative interactions, but we acknowledge some limitations to our study that suggest opportunities for future research. A key methodological consideration was our focus on individual experiences within social settings, which guided our decision to primarily analyze data at the individual level. Despite calculating ICC values that indicated group-level dependency, the limited number of groups (\textit{N} = 8) precluded robust group-level statistical analysis. We did not account for the potential influence of prior relationships within each participant group, as initial impressions of avatars may vary between groups with and without pre-existing relationships. Future research should employ designs with larger numbers of groups to more investigate multilevel effects and distinguish between individual and group-level factors influencing avatar-mediated communication. 

Given technology constraints at the time of our data collection, participants used two separate screens (mobile for FaceTime, laptop for activities shared on Zoom), which may have increased cognitive load despite no reported issues in interviews. Leveraging current advancements in integrated avatar features on single platforms may eliminate this potential confound. Our game-based approach, while in line with common practices  in HCI and avatar research, used brief puzzle-type challenges that may not replicate the complexity or duration of real professional meetings. Research in more naturalistic, longer-duration collaborative contexts and professional settings would provide valuable ecological validity.

While we compared avatar and video representations, future work should include an audio-only condition to determine whether avatars offer unique benefits beyond those provided by removing visual self-presentation entirely. This would provide a more complete picture of the trade-offs among multiple representational affordances of interaction modalities. The requirement for Apple devices potentially introduced selection bias beyond demographics, as Apple users may differ in technological adoption patterns and other relevant characteristics. The cultural homogeneity of our participants (29 of 32 identifying as having Asian heritage), too, may have shaped responses in culturally specific ways. Future studies should recruit participants with more diverse technological and demographic backgrounds to enhance generalizability, given documented cross-cultural differences in self-presentation concerns and collaboration styles. Furthermore, the design characteristics of Memoji avatars (i.e., cartoonish and playful) likely played a role in shaping participants' perceptions and experiences. The observed progression from initial self-consciousness to more comfortable use may partly reflect novelty effects of using avatars in 2D video-based interfaces rather than sustainable benefits, even among participants who were familiar with Memoji. Future research should examine whether the observed benefits of avatars persist over extended use or diminish as the novelty wears off.

\section{Conclusion}

Combining a within-subject experimental design with in-depth group interviews, our study explores how avatars, as a camera-driven representation, affect collaborative interactions in groups working through non-immersive, video-based interfaces. We find that avatar representations significantly enhance self-esteem and video-based collaborative interaction experience compared to traditional video representation. In addition, people experienced key benefits (e.g., increased self-confidence, engaged participation, comfort of physical self, and reduced social tension) but also faced challenges (e.g., emotional expression and interpretation). Our findings support and extend previous research on avatar-mediated communication by providing new empirical insights and implications concerning how avatar use in collaborative interactions impacts self-esteem, collaboration satisfaction, and perceptions of self and others.

\begin{acks}
  We would like to thank Sophie Park, Connie Yang, Jinhan Wu, Cassandra Johnson, Nikta Farahani, and Parniyan Rahbar for their dedicated effort in the ideation and early iterations of the game-based map element design and their assistance in piloting the study. We also thank 20 student researchers in our Fall 2023 directed research group for their assistance in qualitative coding.
\end{acks}

\bibliographystyle{ACM-Reference-Format}
\bibliography{main}


\appendix

\section{Group Interview Protocol} \label{appendix:group-interview}

Prior to starting the group interview, we restated the information outlined in the informed consent, which had been obtained from all participants prior to the start of the study, and addressed any questions participants may have had. The semi-structured interview questions were as follows:
\begin{itemize}
    \item What are your thoughts about using avatars to communicate with your team?
    \item What were some aspects, whether positive or negative, of using avatars during team communication? How/Why?
    \item How did using avatars impact your ability to understand how your team members during the collaboration?
    \item How did the use of avatars influence your team's interactions and contributions during the activities?
    \item In what situations would it feel appropriate and most beneficial to you to use an avatar?
    \item What factors would most influence your decision to use or not use an avatar in a virtual meeting? (Consider factors such as: 1) if others are using avatars; 2) if the avatar could be easily customized before joining the meeting; and 3) if the avatar represented my facial expressions with greater accuracy).
\end{itemize}
We were mindful that some participants might feel less comfortable speaking up spontaneously in a group setting, even in a familiar one. We encouraged individual input by asking follow-up questions such as ``What do you think about that?'' and ``Can you elaborate on your perspective?'' Following the interviews, we provided a debrief to thank participants and address any remaining questions or concerns.

\end{document}